\documentstyle[twoside,fleqn,espcrc2,epsf]{article}

\newcommand{\be}{\begin{equation}}
\newcommand{\ee}{\end{equation}}
\newcommand{\ba}{\begin{eqnarray}}
\newcommand{\ea}{\end{eqnarray}}
\def\simm#1{\mathop{\vtop{\ialign{##\crcr
	$\hfil\displaystyle{#1}\hfil$\crcr\noalign{\kern0.5pt\nointerlineskip}
	$\sim$\crcr\noalign{\kern0.5pt}}}}\limits}
\def\mpsq{m_{\pi}^2}
\def\Ninv{N_{\rm inv}}
\def\nf{N_F}
\def\nt{N_t}
\def\kc{K_C}
\def\kt{K_T}
\def\kd{K_D}
\def\bct{\beta_{CT}}

\title{Quantum chromodynamics with various number of flavors
\thanks{Talk presented by K. Kanaya 
at {\it Lattice 93}, Dallas, USA.}
}

\author{Y. Iwasaki\rlap,\address{Institute of Physics, University of Tsukuba,
        Ibaraki 305, Japan} 
        K. Kanaya\rlap,$^{\rm a}$
        S. Sakai\address{Faculty of Education, Yamagata University, 
        Yamagata 990, Japan}
        and 
        T. Yoshi{\' e}$^{\rm a}$
       }       
\begin{document}

\begin{abstract}
The phase structure of QCD with various number of flavors is studied 
for Wilson quarks. For the case of $\nf=3$ we find that the finite 
temperature deconfining transition is of first order in the chiral 
limit on an $\nt=4$ lattice. Together with our previous results that 
the deconfining transition in the chiral limit is continuous for 
$\nf=2$ and is first order for $\nf=6$, the order of the transition 
is found to be consistent with a prediction of universality. 
The case of $SU(2)$ QCD is also studied in the strong coupling limit 
and the phase structure is found to be quite similar to the case of 
$SU(3)$: There exists a critical number of flavors $\nf^*$ and for 
$\nf \geq \nf^*$ the confinement is broken even in the strong coupling 
limit for light quarks. $\nf^*=3$ corresponding to 7 for $SU(3)$. 
\end{abstract}

\maketitle

\section{INTRODUCTION}
The deconfining transition of QCD is the last phase 
transition which our Universe has enjoyed in the past. 
To study the nature of this transition 
it is essential to include dynamical quarks. 
In this paper we study the phase structure of finite temperature 
lattice QCD with degenerate $\nf$ Wilson quarks. 

Our model has 4 parameters: 
$\nf$ (the number of flavors), $\beta$ (gauge coupling), $K$ (hopping
parameter) and $\nt$ (lattice size in the temporal direction).
For $\nt=4$ -- 8 we use a spatial $8^2 \times 10$ lattice 
(the lattice size 10 is doubled for hadrons), 
while for $\nt=18$ we use $18^2 \times 24$.
The method of simulation is similar to that described in 
\cite{ourLat92,ourPRL92}. 

\begin{figure}[tb]
\epsfxsize=6.5cm \epsfbox{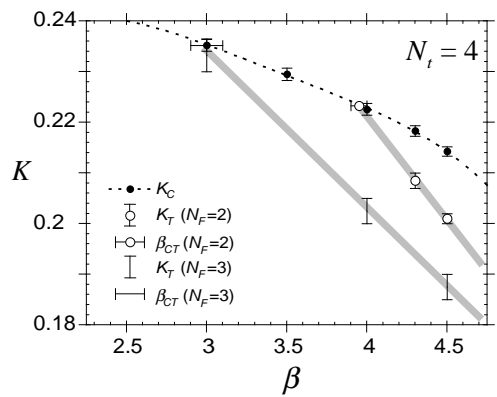}
\vspace{-0.9cm}
\caption{Phase diagram for $\nf=2$ and 3 on an $\nt=4$ lattice.}
\label{fig:F2F3PDiag}
\end{figure}

\subsection{$\nf$ = 2 and 6}
In a previous paper \cite{ourLat92} we studied the 
deconfining transition $\kt$ for the cases of $\nf=2$ and 6 
on an $\nt=4$ lattice. 
Fig.~\ref{fig:F2F3PDiag} shows our result of the phase diagram 
for $\nf=2$ near the chiral limit $\kc$.
(The result for $\nf=3$ will be discussed later.)
$\kt$ is identified by a sudden change in the behavior of physical 
observables such as plaquette and $\mpsq$. 
$\kc$ is determined by a linear extrapolation (in $1/K$) 
of $\mpsq$ in the confining region on the lattice with the same $\nt$.
We find that $\nt$- (as well as $\nf$-) dependence of $\kc$ is small 
for $\beta \,\simm{<}\, 4.5$ 
compared with the difference of $\kc$ determined by $\mpsq$ and $m_q$;
the difference is a result of the O($a$) corrections concerning 
the chiral behavior of Wilson fermions. 
The location of the crossing point $\bct$ is identified 
by monitoring $\Ninv$, the number of CG iterations to invert the quark 
matrix, just on the $\kc$-line \cite{ourLat92,ourPRL92}. 
This can be done because of an empirical law \cite{ourPRL91} that $\Ninv$ 
becomes very large in the confining phase when we get close to $\kc$, 
while $\Ninv$ remains small in the deconfining phase 
even at $\kc$. 

The nature of the transition at $\bct$ can be studied by 
measuring observables on the $\kc$-line \cite{ourLat92}.
We find that, when we decrease $\beta$ toward $\bct$,
$\mpsq$ decreases linearly and is consistent with zero at $\bct$.
This implies that the transition is continuous (second order
or crossover), in accord with the result of an universality 
argument by Rajogopal and Wilczek who predict 
the second order transition for $\nf=2$ 
\cite{RajWilcz}. 

This result is in clear contrast with the case of $\nf=6$ where
we find that $\mpsq$ is large near and above $\bct \simeq 0.3$. 
We also find a two-state signal in the time development
of $\Ninv$ and other observables at $\beta=0.3$ \cite{ourLat92}.
The first order transition observed is again consistent with a prediction 
of an universality argument stating that the transition is of 
first order for $\nf \geq 3$ \cite{PisWilcz}.

\begin{figure}[tb]
\epsfxsize=6.5cm \epsfbox{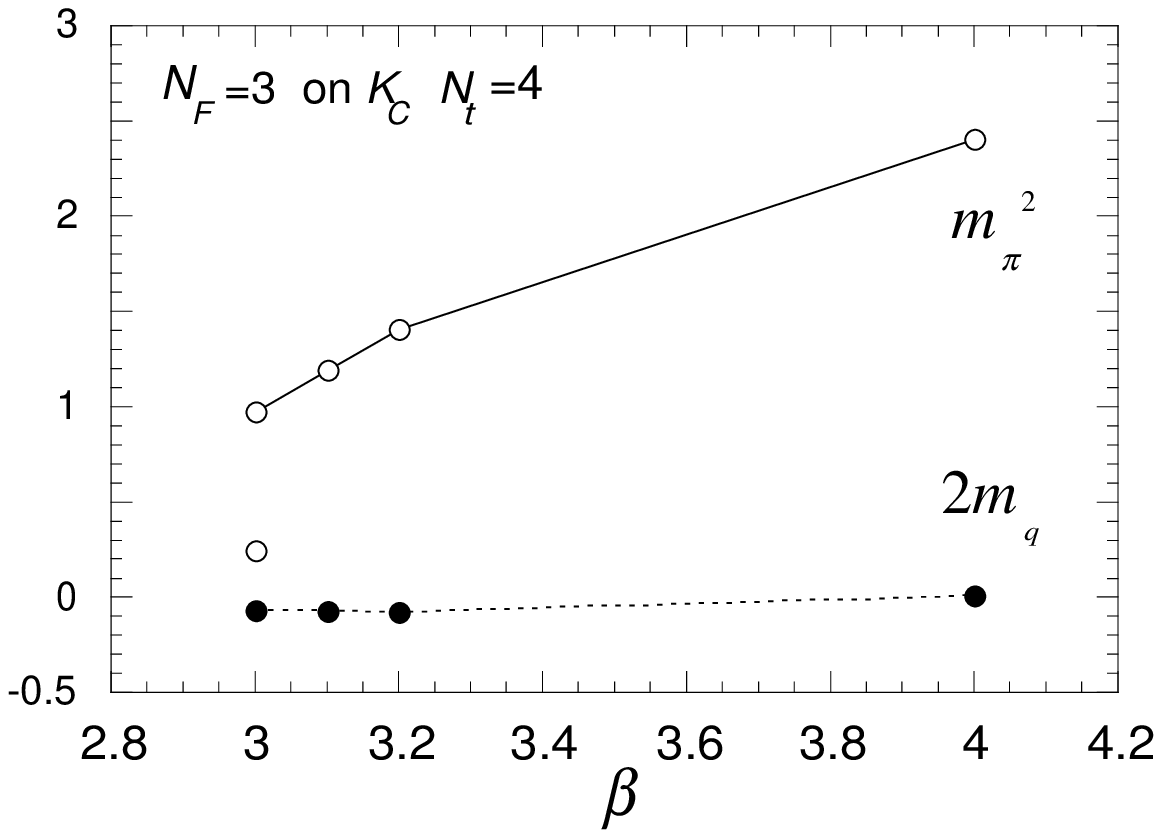}
\epsfxsize=6.5cm \epsfbox{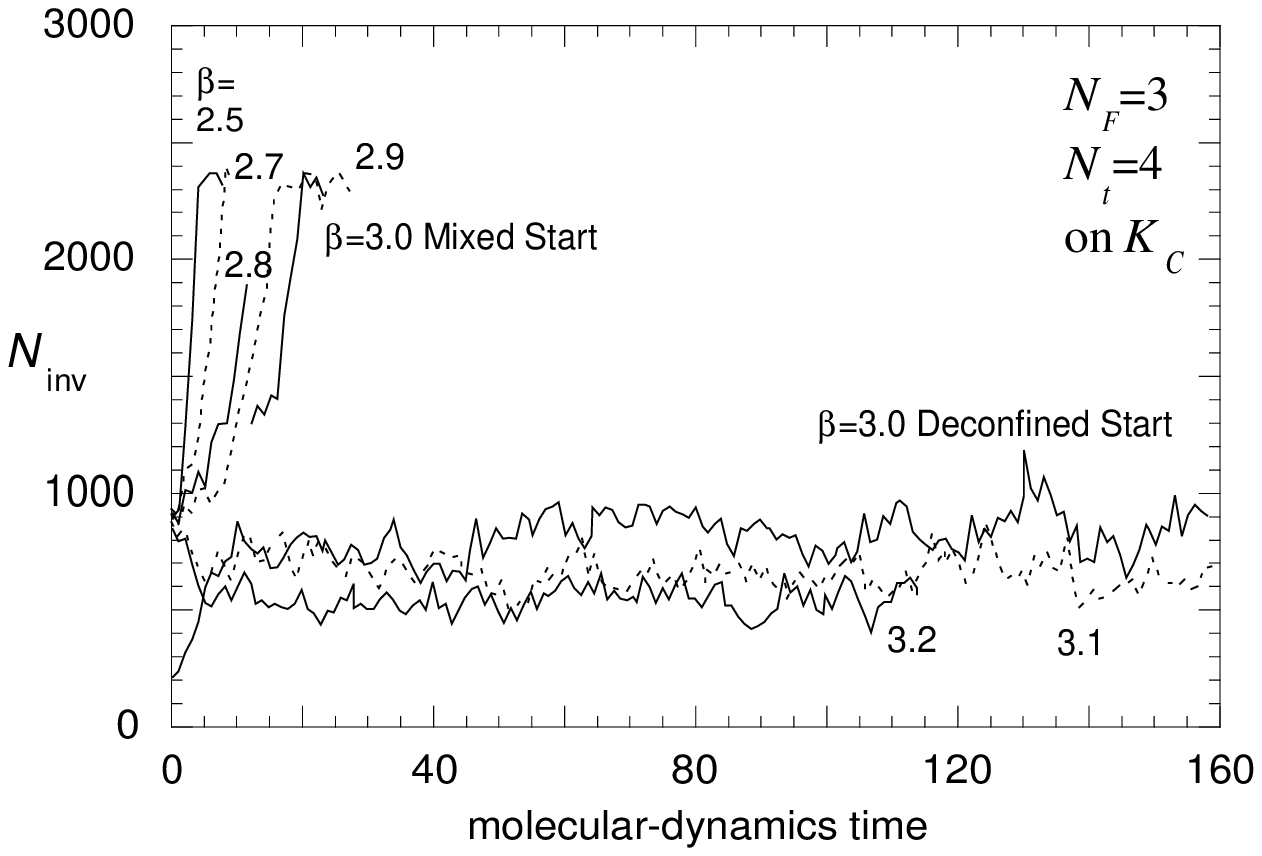}
\vspace{-0.9cm}
\caption{(a) $m_{\pi}^2$ and $2m_q$ 
and (b) $\Ninv$ history for $\nf=3$ on $\kc$-line. 
Jackknife errors for (a) are smaller than the symbols.}
\label{fig:F3Kc}
\end{figure}

\section{$\nf$ = 3}

Because the effective $\nf$ at the deconfining transition temperature 
in the real world is between 2 and 3, it is important to extend the study 
to the case of 3 flavors. 
Performing simulations 
we find the phase diagram shown in Fig.~\ref{fig:F2F3PDiag}.
As expected, $\kt$ shifts to smaller $\beta$ with increasing 
$\nf$ from 2 to 3. 
Fig.~\ref{fig:F3Kc} (a) is the result for $\mpsq$ on the $\kc$ line. 
We find that $\mpsq$ is large for $\beta \,\simm{>}\, \bct \simeq 3.0$. 
We also find a two-state signal at $\beta=3.0$ 
(Fig.~\ref{fig:F3Kc} (b)).
We therefore conclude that the deconfining transition is of first order 
for $\nf=3$ in the chiral limit, 
which is in accord with the prediction of universality 
by Pisarski and Wilczek \cite{PisWilcz}.

\section{MANY FLAVORS}

\subsection{$SU(3)$ case}

Now let us consider the case of much larger $\nf$.
In our previous paper we have shown that, when $\nf \geq 7$, light 
quarks are not confining even in the strong coupling limit\cite{ourPRL92}. 
Extending the study to finite $\beta$ 
we find the following phase structure for $\nf \geq 7$\cite{ourLat92}:
We have a $\nt$-independent deconfining transition line $\kd$,
which starts from a finite $K$ at $\beta=0$ and extends to larger $\beta$.
In addition, we have a $\nt$-dependent deconfining
transition/crossover line $\kt$,
which reaches the $\kd$-line without crossing the $\kc$-line.

To understand this phase structure, let us recall the property of the 
perturbative $\beta$ function. 
Asymptotic freedom is broken for $\nf \geq 17$ because the first coefficient 
of the $\beta$ function becomes positive. 
We note that the second coefficient changes its sign for $\nf \geq 9$. 
This suggests the existence of an infrared fixed point (IRFP) for 
$16 \geq \nf \geq \nf^*$ ($\nf^*=9$ up to 2-loop perturbation theory). 
With an IRFP 
correlation functions scale with non-canonical powers at large distances 
and therefore there are no hadronic particles\cite{BanksZaks}. 
Our MC result \cite{ourLat92,ourPRL92} implies that $\nf^*=7$.

\begin{figure}[t]
\epsfxsize=6.4cm \epsfbox{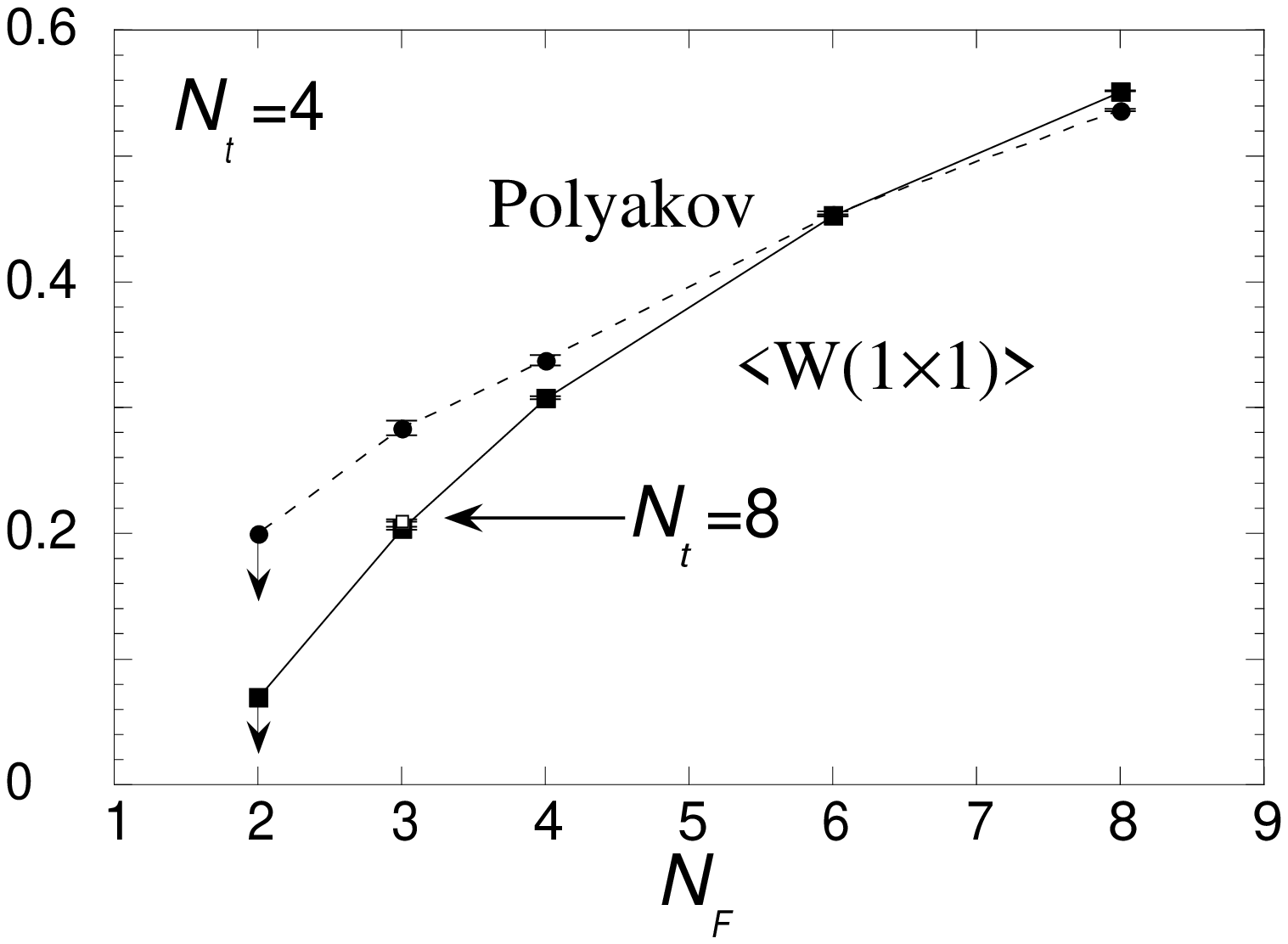}
\epsfxsize=6.4cm \epsfbox{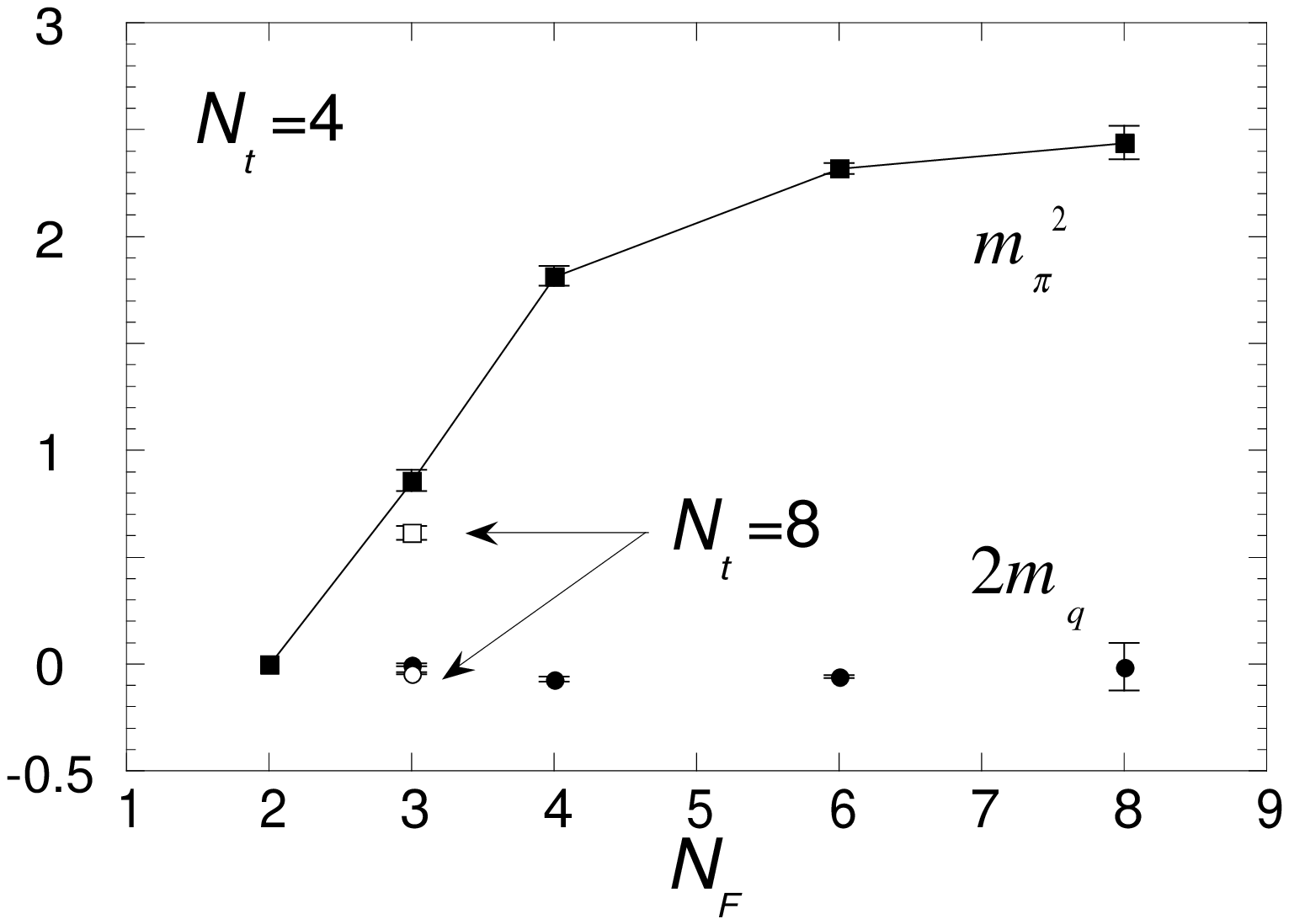}
\vspace{-1.0cm}
\caption{Observables in $SU(2)$ QCD for $K=0.25$ at $\beta=0$ 
as a function of $\nf$: (a) plaquette and Polyakov 
loop and (b) $\mpsq$ and $2m_q$.}
\label{fig:SU2Kc}
\end{figure}

\subsection{$SU(2)$ case}

As an extension of the $SU(3)$ case, we study an $SU(2)$ QCD 
in the strong coupling limit. 
We note that the $SU(2)$ $\beta$-function has the same characteristics 
as $SU(3)$ 
except for that typical numbers for $\nf$ are generally smaller 
than those for $SU(3)$:
With $SU(2)$, asymptotic freedom is lost when $\nf \geq 11$ 
and 2-loop result for $\nf^*$ is 6. 
This is natural because the confining force is weaker for $SU(2)$.
We may therefore expect that $\nf^*$ is smaller than the case of $SU(3)$.

Performing a simulation for $\nf=8$ and $\nt=4$ 
we find that the deconfining transition appears at $K = 0.2$ -- 0.21 
and that the chiral limit at $K=0.25$ is in the deconfining phase. 
Decreasing $\nf$ gradually with fixing $K=\kc$, we find that the chiral 
limit remains in the deconfining phase down to $\nf=3$. 
When we further decrease $\nf$ to 2, $\Ninv$ shows a rapid increase 
with the molecular-dynamics time and finally exceeds $10\,000$ 
in clear contrast with small numbers of O($10^2$) for $\nf \geq 3$. 
The chiral limit is therefore in the confining phase for 
$\nf=2$ \cite{ourPRL92}. 
Fig.~\ref{fig:SU2Kc} shows our results for physical observables 
for $K=\kc$. 
To study the $\nt$ dependence of the transition, we simulate on 
an $\nt=8$ lattice for the critical case of $\nf=3$. 
The stability of the result confirms that the deconfining transition 
we observe is a bulk transition. 
According to our experience with $SU(3)$ QCD, we expect that these 
properties in the strong coupling limit hold also for small $\beta$. 
We therefore conclude that $\nf^*=3$ for $SU(2)$ QCD. 

\section{CONCLUSION}
We studied the phase structure of QCD with dynamical Wilson quarks.
Extending our previous study, we found that the finite temperature 
deconfining transition for $\nf=3$ is of first order. 
We also studied an $SU(2)$ QCD and found that $\nf$-dependence of 
$SU(2)$ QCD is quite similar to that of $SU(3)$, but now $\nf^*$, 
above which confinement is lost for light quarks, 
is 3 instead of 7 for $SU(3)$. 

We have studied the phase structure of QCD mainly on an $\nt=4$ lattice. 
To study the implication to the continuum limit, 
the system should be investigated also at larger $\beta$ on a larger lattice 
\cite{MILC,Columbia93}. 
We are extending our study in this direction. 

The simulations are performed with HITAC S820/80 at KEK and with QCDPAX. 
We thank members of KEK for their support
and the other members of QCDPAX collaboration for their help.
This project is in part supported by the Grant-in-Aid
of Ministry of Education, Science and Culture
(No.62060001 and No.02402003).


\end{document}